\documentclass{article}[12pt]
\usepackage{everypage}
\usepackage{authblk}
\usepackage{graphicx}
\usepackage{etex}
\usepackage{float,verbatim}
\usepackage{color,ulem}
\usepackage{amssymb,amsmath}
\RequirePackage{xspace}
\def\be{\begin{equation}}
\def\ee{\end{equation}}
\def\bee{\begin{eqnarray}}
\def\eee{\end{eqnarray}}

\begin{document}

\title{ Why the interpretation of ``Measuring propagation speed of Coulomb fields" stands}

\author{R. de Sangro, G. Finocchiaro, P. Patteri, M. Piccolo, G. Pizzella\\
\small{ Istituto Nazionale di Fisica Nucleare, Laboratori Nazionali di Frascati}}

\maketitle

\abstract { The experimental findings  reported in our original paper \cite {risultati}  have been criticized in ref.\cite{shabad}.
We believe that the arguments brought  in  ref. \cite{shabad} are not correct and we show evidence for this.} 
\section{Introduction}
In ref. \cite{shabad}, our measurements \cite{risultati} have been criticized: the authors imply that the responses of our sensors are not caused by the Coulomb field carried by the electron beams, but  are rather due to  electromagnetic radiation, generated in the last magnetic bend the electron beam undergoes entering the experimental hall.
The detail of their main points are:
\begin{enumerate}
\item Inconsistencies  in our longitudinal timing  measurements.
\item Sloping level of background vs.transverse distance for the beam dump measurements.
\item Sensor signals due to synchrotron radiation in the last bend of the beam line.
\end{enumerate} 
In the following we will address the stated points separately.

\section {Longitudinal time measurements}
It is clearly stated in our paper \cite{risultati} (fig.\;13) that no timing dependence on the transverse position of our sensors was measured, and  that the correlation
 timing-longitudinal position was the one pertaining to an electron beam moving with $\gamma \approx 1000$ in the experimental hall. 
Out of the six measurements reported in table 1 of ref. \cite{risultati}, there is just one, $-2.8 \sigma$ away from the above mentioned hypothesis, and is the one that has been speculated upon  in ref. \cite{shabad}. The authors point out that the measurement might infringe the speed of light barrier.
To put things into perspective the 2.8 $\sigma$  discrepancy corresponds to less than 150\,psec in time or less than 4\,cm in space.

We stress  that, considering the entire set of data  reported in the aforementioned table, one readily sees that timing for sensor A5 tend to be {\it early} while timing for sensor A6 tend to be {\it late}.
Averaging the timing for A5 and A6, in fact, wipes out completely the fluctuation, and strongly suggests that there  be a systematic effect due to a less than perfect flatness of the experimental hall floor. A 35\,mrad angle would grant the effect seen in table 1 of ref. \cite{risultati}.

We summarize the time distance correlations  in Table \ref{time}, where the sensors' A5,A6 {\it average} time  
obtained at the three different longitudinal positions are shown.

\begin{table}[hbtp]
\begin{center}
\caption{Timing measurements. The expected differences  are calculated for
500 MeV electrons. The agreement between the calculated  and the experimental values  is more than satisfactory. }
\vspace{3 mm}
\begin{tabular}{|c|cc|}
\hline
longitudinal distances&expected&experimental\\
between two sensors [cm]&[ns]&$\overline{A5,A6}$ [ns]\\
\hline
$(552.5-329.5) ~223.0\pm1.5$&$7.43\pm0.05$&$7.40\pm0.06$\\
$(552.5-172.0) ~380.5\pm1.5$&$12.68\pm0.05$&$12.73\pm0.09$\\
$(329.5-172.0) ~157.5\pm1.5$&$5.19\pm0.05$&$5.19\pm0.07$\\
\hline
\end{tabular}
\label{time}
\end{center}
\end{table}

\section{Sloping background in the beam-dump measurements.}
No quantitative statement was made in \cite{shabad} for this effect. In order to have a quantitative understanding of the data, we fit the four point  reported in fig 15 of ref \cite{risultati} either with a constant or with a first order polinomial.
A comparison between the $\chi^{2}\over DOF$  for the two hypotheses might give an insight, in relative terms, as to which one is more suited to represent the data.

\begin{table}[hbtp]
\begin{center}
\caption{Fit results: flat vs sloping background. Beam dump measurements.}
\vspace{3 mm}
\begin{tabular}{|c|c|c|c| }
\hline
& P0 & P1 & $\$chi^{2}\over D.O.F. $\\
\hline
flat &0.047$\pm$ 0.006& - &2.9\\
\hline
sloping&0.038$\pm$ 0.0098&0.004 $\pm$ 0.0034&3.6\\
\hline
\end{tabular}
\label{fitcomp}
\end{center}
\end{table}

While it is clear that neither  hypothesis fits the data very well, the zeroth order polynomial shows a better agreement to the experimental points; it is worth noticing that the linear function's fitted slope is really ill determined with a relative error close to 80\%.

\section {Synchrotron radiation out of the last bend}

Here too we do not have any quantitative statement in ref.\,\cite{shabad}, so we evaluate here the electromagnetic power  released in the last bend and compared it  with the power of our sensors' signal.

The beam line into the experimental hall has a 45 degrees  bend at 1.72\,m curvature radius: this translates into a synchrotron radiation (total) power of $\approx 510 \times 10^{-3}$ W for a typical pulse of $10^{8}$ electrons at 500 MeV.
The critical frequency for the bend is  $\omega_{c} \approx 3\times 10^{17}$ Hz.

We did evaluate the synchrotron radiation power hitting our sensor either with no angular cut or by means of a full Monte Carlo simulation \cite{francesco} taking into account the angular distribution of the emitted radiation at the position of the most exposed detector (radial detector(s)  at 5 cm transverse distance and 92 cm. longitudinal distance)  
As clearly stated in   ref.\,\cite{risultati}, our sensors have a cut-off frequency of $\approx 250$ MHz.

The results are summarized in table 3.

\begin{table}[hbtp]
\begin{center}
\caption{Syncrotron radiation summary}
\vspace{3 mm}
\begin{tabular}{|c|c|c| c|}
\hline
&  Total power &  Power below 250 MHz & Aver. sens. pulse power \\
\hline
No angular cut & 510 $\times10^{-3}$W & 2{$\times10^{-11}$} W & 1 $\times 10^{-6}$ W\\
\hline
Full Monte Carlo & 106 $\times 10^{-3}$W &  4.$\times10^{-12}$ W & 1 $\times10^{-6}$ W\\
\hline
\end{tabular}
\label{sync}
\end{center}
\end{table}

The amount of power synchrotron radiation conveys onto our sensors is at least  $\approx 50000$ times smaller than the one pertaining to our  measured pulse amplitude.

%{\small\date}

\end{document}